\documentclass[useAMS,usenatbib]{mn2e}

\usepackage{graphicx}
\usepackage{epsfig}

\title[]{Jeans Instability of  Palomar 5's Tidal Tail}
\author[]{
Alice C. Quillen \&  
Justin Comparetta \\
{Department of Physics and Astronomy, University of Rochester, Rochester, NY 14627, USA;}\\
{  aquillen@pas.rochester.edu; jcompare@pas.rochester.edu}  \\
}

\begin{document}
\label{firstpage}
\maketitle

\begin{abstract}

Tidal tails composed of stars should be unstable to the Jeans instability and this can cause them to look like beads on a string.   The Jeans wavelength and tail diameter determine the wavelength and growth rate of the fastest growing unstable mode.  Consequently the distance along the tail to the first clump and spacing between clumps can be used to estimate the mass density in the tail and its longitudinal velocity dispersion.    Clumps in the tidal tails of the globular cluster Palomar 5 could be due to Jeans instability.  We find that their spacing is consistent with the fastest growing mode if the velocity dispersion in the tail is similar to that in the cluster itself.   While all tidal tails should exhibit gravitational instability, we find that clusters or galaxies with low concentration parameters are most likely to exhibit short wavelength rapidly growing Jeans modes in their tidal tails.

\end{abstract}

\section{Introduction}
 
A thin self-gravitating cylinder of gas can clump into a series of bead like structures (e.g., \citealt{chandra53,ostriker64,elmegreen79,elmegreen94}).
This model has been use to propose explanations for clumps of star formation and star clusters along spiral arms \citep{elmegreen83} and in tidal tails of galaxies \citep{elmegreen96,duc00,smith10}.  The sausage or varicose instability is a rough analogy to the Plateau--Rayleigh  instability where surface tension causes a stream of water to become a series of droplets.
A cylinder of stars can also contract into a series of clumps.  In 
this case the stability has been called the Jeans instability \citep{fridman84} even though this term is more often used to describe the growth and decay of plane waves through a homogeneous stellar medium.  The instability for a rotating stellar cylinder has
been investigated by \citet{fridman84} (see their Chapter 2).  In the limit of infinitely slow rotation their analysis can be used for the non-rotating cylinder.

Nearly equally spaced clumps observed in the tidal debris of Palomar 5 \citep{oden01,oden02} has not been interpreted in terms of the Jean instability but 
instead primarily in terms of oscillations in the cluster \citep{gnedin99} 
caused by a previous passage through the Galactic disk \citep{oden02,dehnen04}.
Alternative explanations for structure in tidal tails 
include the effect of structure in the dark matter halo, as explored by \citet{mayer02}
and variations in density along the tail caused by epicyclic motions in the stellar orbits 
\citep{kupper08,just09,kupper10}.

Here we consider the possibility that clumps in a tidal streams could be caused by gravitational instability.

We review the properties of the Palomar 5 cluster and tail system.
Its distance is estimated at 23.2 kpc \citep{harris96}.
The velocity dispersion in the tail is small, measured at 2-4 km/s \citep{oden09}.
The dispersion in the cluster itself is small,  $\sim 0.3$ km/s (correcting for binaries \citealt{oden02}), and the cluster is suspected to have low concentration parameter,
$c_t$, in the range 0.4-0.8 \citep{oden02,dehnen04}.
The width of the tail is 120~pc and the tidal radius of the cluster half of this \citep{oden02,oden03}.  The mass of the cluster is about 5000 $M_\odot$ \citep{oden02}
and that in the tail extending 10$^\circ$ exceeds the mass in the cluster 
by a factor of a few \citep{oden02,oden03}.   The tail has been detected
up to 22$^\circ$ from the cluster \citep{grillmair06}. The linear mass density in the two tails as a function of distance along the tail drops slowly as a function of distance from the cluster \citep{oden03} as would be expected from a tidal evolution model \citep{johnston99,dehnen04}.

Clumps are most easily seen on the southern side of the tidal tail with distance to
the first clump about 1.7 degrees from the cluster center and that
between the first and second clump about 2 degrees (using Figure 14 by \citealt{oden03} that also shows that extinction cannot account for the density variations).
At a distance of 23.2 kpc, 2 degrees corresponds to 800~pc and we can treat
this as the wavelength of a possible growing unstable mode.  It is useful to 
take the ratio of the wavenumber to the tidal tail radius, $r_0$, (diameter/2).  
We estimate $k r_0 \sim 0.5$ using $r_0 \sim 60$~pc.


\section{Jeans instability of a stellar cylinder}
  
We briefly review the gravitational instability of a stellar cylinder to varicose or sausage like compressive modes of oscillation.  The dispersion relation is derived more
rigorously by \citet{fridman84} (in their Chapter 2) and covering the case of a homogenous rotating stellar cylinder.
We can adopt the simplest assumption of a constant linear mass density, $\mu_0$, but a deformed boundary of radius $R(z)$ and a corresponding interior density, $\rho$, that is independent of radius interior to $R$.  We are working in cylindrical coordinates with $z$ oriented along the cylinder.
 To first order in a perturbation amplitude $a$
\begin{eqnarray}
\rho(z) &=& \rho_0 \left( 1 + a \cos kz \right) \nonumber \\
R(z) &=& r_0 \left( 1 - {a \over 2} \cos kz \right)
\label{eqn:boundary}
\end{eqnarray}
Here we have assumed that our perturbation amplitude, $a$, is small.
 The relation between the perturbation amplitude
in radius $R$ and density $\rho$ is determined to first order by the assumption that the 
linear mass density does not depend on $z$ and $\mu_0 = \rho_0 \pi r_0^2$.

To estimate the perturbation to the gravitational potential we can consider the potential perturbation caused by a massive wire with 
mass density $\rho(r,z) = \delta(r=0)(\mu_0 +  \mu_1 \cos(kz))$.  The part of the gravitational potential fluctuating with $z$ at a distance $r$ from the wire and at $z$ is
\begin{eqnarray}
\Phi(r,z) 
&=& -\int_{-\infty}^\infty G \mu_1 { \cos(kz') \over \sqrt{r^2 + (z-z')^2 }} dz' \nonumber \\
&=&  -2 G \mu_1 \cos(kz) K_0(k r)
\label{eqn:bes} 
\end{eqnarray}
where $K_0$ is a modified Bessel function of the second kind.  

To estimate the gravitational potential perturbation  we must integrate the Bessel function over radius interior to the cylinder boundary.
For $k r < 1$ the asymptotic limit $K_0( k r) \to - \ln (k r/2) - 0.5772..$
where the constant is the EulerÐ-Mascheroni constant. 
Integrating equation \ref{eqn:bes} out to our boundary $R$ (given in equation \ref{eqn:boundary}) the potential perturbation
\begin{equation}
\Phi_1(z,r=0) \approx \int_0^R  4 \pi G \rho_0 a \cos kz 
\left[\ln \left({k r\over2}\right)+ 0.5572\right] r dr
\end{equation}
Deviations on the boundary only contribute to second order so
\begin{equation}
\Phi_1(z) \approx - 2 G \mu_0 a |1 - \ln k r_0| \cos kz
\label{eqn:grav}
\end{equation}
where we have used a constant of 1 inside the absolute value as it is a more accurate
match to the integral of the Bessel function than the constant 0.62 given by the asymptotic limit.
To first order in $a$ the same relation would have been estimated if we had assumed
no boundary deformation but a longitudinal variation in linear density $\mu$.  This situation corresponds to longitudinal modes rather than sausage or varicose modes
(e.g., \citealt{ostriker64}).

We now consider density and potential perturbations that
are proportional to $e^{i(kz - \omega t)}$.
We assume a phase space stellar distribution function in the cylinder 
\begin{equation} 
f(z,v,t) = \rho_0 \left[1 + a e^{i(\omega t - k z)} \right] f_a(v)
\label{eqn:ff}
\end{equation}  
consistent with our density perturbation (equation \ref{eqn:boundary}).
Here $f_a(v)$ is a Gaussian velocity distribution and we need only consider the distribution as a function of the $z$ velocity component,
\begin{equation}
f_a(v) = {1 \over \sqrt{2 \pi \sigma_*^2}} \exp\left({-v^2 \over 2 \sigma_*^2}\right) 
\end{equation}
where $\sigma_*$ is the stellar velocity dispersion in the $z$ direction.

We use a linearized version of the collisionless Boltzmann equation,
\begin{equation}
\left[ {\partial  \over \partial t} +  v {\partial  \over \partial z} \right] f_1- {\partial \Phi_1 \over \partial z} {\partial f_0 \over \partial v} = 0
\end{equation}
where we have denoted the unperturbed and first order perturbation to the distribution function as  $f_0$ and $f_1$  (see equation \ref{eqn:ff}). Using equation \ref{eqn:grav} for the potential perturbation we find
\begin{equation}
1 +  {2 G \mu_0 |1-\ln (k r_0)| \over \sigma_*^2} \int  dv  {kv \over \sqrt{2 \pi \sigma_*^2}} {\exp\left({-v^2 \over 2 \sigma_*^2}\right) \over \omega - kv }= 0
\end{equation}
valid for $k r_0 \la1$.
We can rewrite this as
\begin{equation}
1 - q^{-1} |1-\ln k r_0| \int_{-\infty}^\infty {1 \over \sqrt{\pi}}  {e^{-u^2} u du  \over u-Z } =0
\label{eqn:zeq}
\end{equation}
where $Z = \omega/ (\sqrt{2} k\sigma_*)$
and 
\begin{equation}
q \equiv {\sigma_*^2 \over 2 G \mu_0}  = {2 \over (k_J r_0)^2}
\label{eqn:qdef}
\end{equation}
where $k_J \equiv \sqrt{4 \pi G \rho_0 \over c_s^2}$ is the Jeans wavenumber
and $k_J = 2\pi/\lambda_J$ with $\lambda_J$ the Jeans wavelength.

The value $q$ is analogous to the Toomre Q instability parameter.
The dispersion relation is similar to that for the Jeans instability for a plane wave propagating through a homogenous stellar medium.  
The above dispersion relation that we have estimated roughly from the potential perturbation at the center of the cylinder is consistent with that in equation (14) more rigorously derived by \citet{fridman84} for the homogenous rotating cylinder.  

Assume that $\omega = i \gamma$ leading to unstable solutions.
Then we can integrate equation \ref{eqn:zeq}  finding 
\begin{equation}
1 - q^{-1} |1-\ln k r_0| \left[\sqrt{\pi} e^{\gamma'^2} \gamma' \left({\rm erf} \gamma' - 1\right) + 1 \right]=0
\end{equation}
where $\gamma' = \gamma /(\sqrt{2} k \sigma_*)$.
This dispersion relation  implies that a ``long cylinder is unstable 
to any large but finite thermal scatter of particles in longitudinal velocities'' (\citealt{fridman84} just before their equation 21; also see their Figure 10 showing a numerically measured stability boundary for the non-rotating cylinder with
numerical calculations done by S. M. Churilov).  

The above equation gives a relation between growth rate and wavelength.
As $\gamma'$ depends on $k$ the growth rate is most easily discussed in terms
of $\gamma' k r_0 = \gamma r_0 /\sigma_*$ or the growth rate in units
of the tidal tail's radius and velocity dispersion.
Interestingly the  growth rate seems to be of order $\sigma_*/r_0$
rather than  of order $\sqrt{G\mu_0}k$.  However if $q\sim 1$ and $k r_0\sim 1$
then the two growth rates are similar.

\begin{figure}[h]
\begin{center}
\includegraphics[scale=0.65, angle=0]{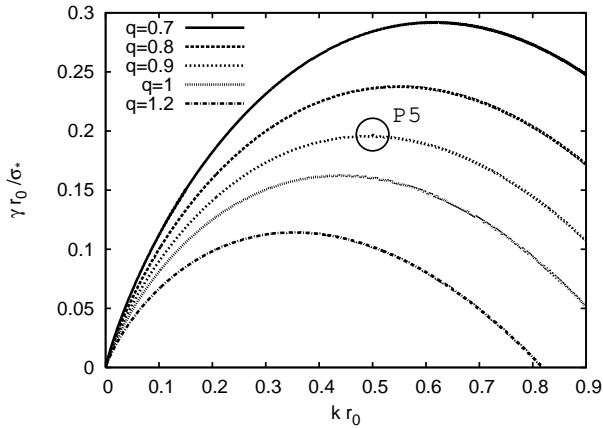}
\caption{Growth rate vs wavelength for 5 different values of $q$.
The wavelength of maximum growth rate depends on $q$ with smaller wavelength
(larger wavenumber $k$) having a faster growth rate if $q$ is smaller.
The $x$ axis is the wavenumber times the tidal tail radius.
The $y$ axis is the growth rate in units of $r_0/\sigma_*$.
For Palomar 5 we have estimated from the distance between
clumps that the fastest growing mode has $k r_0 \sim 0.5$
corresponding to $q \sim 0.9$ and a growth rate of $\gamma r_0/\sigma_* \sim 0.2$.
The estimated growth rate and wavenumber of the fastest growing mode
for Palomar 5's tails is shown with a solid circle in the middle of the Figure.
\label{fig:sofun}
}
\label{fig:fig1}
\end{center}
\end{figure}

In Figure \ref{fig:sofun} we show growth rate versus wavenumber for different values
of $q$.  We see that at each $q$ there is a mode that is maximally unstable or grows fastest.
In this figure we have plotted  $\gamma r_0/\sigma_*$ versus $k r_0$ so that the 
growth rate (shown on the $y$ axis) is independent of the wavenumber.
At larger $k$, the growth rate drops to zero and there is no longer instability.
However for any $q$, even those greater than 1, there are wavenumbers that are unstable.    As $q$ increases these wavenumbers correspond to increasingly long wavelengths and the growth rate becomes slower and slower.   For $q<1$, larger $k$ wavenumber (or shorter wavelength) modes can grow.    

For sufficiently small $q$ we must keep in mind that
we estimated the gravitational potential for $k r_0 \la 1$.   For $k r_0 >1$ (corresponding to wavelengths shorter than the cylinder diameter)
the potential perturbation should be proportional to $k^{-2}$ and we should recover the Jeans instability for plane waves.  
The condition $q <1$ is equivalent to the Jeans wavelength being
shorter than about twice the diameter of the stream (see equation \ref{eqn:qdef}).
For the Jeans instability of plane waves, the larger the wavelength, the faster
the growth rate. Consequently when $q\ll1$ we expect that the fastest growing wavelength will be of order the one that fits within the diameter of the tail, at which point the dispersion relation deviates from that estimated for a plane wave.  

\section{Application to Palomar 5}

In the introduction we estimated from the distance
between the clumps on the southern tail of Palomar 5 that the fastest
growing mode has $k r_0 \sim 0.5$.
Using Figure 1  we estimate that $q \sim 0.9$ in Palomar 5's tail as that
value gives the correct wavelength for the fastest growing mode.
As $q$ depends on both linear mass density and longitudinal velocity dispersion we
have a relation between these two quantities.  We first review estimates
of the linear mass density based on observations and then we place constraints
on both these quantities.

The tidal tail linear mass density for a cluster that is losing mass
at a rate $\dot M$ is 
\begin{equation}
\mu={dM \over dz} = \dot M {dt \over dz}
\label{eqn:mu}
\end{equation}
where $dz\over dt$ is the rate that stars are drifting away from the cluster
and $z$ gives distances along the tail.

At a radius $R$ from the Galaxy center,
\begin{equation}
{d z \over dt} \approx {d \Omega \over dR} R r_t
\label{eqn:dzdt}
\end{equation}
where $\Omega(R)$ is the angular rotation rate of a particle 
in a circular orbit in the galaxy radius $R$.  
This is approximately consistent with the more precise semi-analytical 
model by \citet{johnston99}.
For a galaxy with a flat rotation curve
${d \Omega \over dR} \sim {v_c \over R^2}$
so 
\begin{equation}
{d z \over dt } \approx \Omega r_t. 
\label{eqn:dzdt2}
\end{equation}
For a tidal radius of 60~pc, a galactic rotational velocity of $v_c = 220$~km/s
and distance of 23.2~kpc, the drift rate in Palomar 5's tidal tail
is $dz/dt \approx 0.6$pc~Myr$^{-1}$. 
The above estimate is only approximate as a better estimate would take 
into account the cluster orbit and adopt a more realistic Galactic potential 
(e.g., as explored in the appendix by \citealt{oden03}; also see \citealt{johnston99,just09,kupper10}).

Using the estimated drift rate of 0.6 pc~Myr$^{-1}$  the  mass loss rate of $\dot M \sim 5 M_\odot$ Myr$^{-1}$ estimated by \citet{oden09}
corresponds to a linear density of $\mu \sim 8 M_\odot$~pc$^{-1}$.
This is three times lower than a linear mass density estimated from stellar number counts, with the number counts in
the tails over 10 degrees exceeding that in the cluster by a factor of about 2 and the estimated total cluster mass of $\sim 5000 M_\odot$ \citep{oden02}. 
 We note that mass segregation may affect estimates of 
 the tail mass density based on  star counts  \citep{koch04}.

We can write for $q$
\begin{equation}
q \equiv {\sigma_*^2 \over 2 G \mu} \approx 1
\left({\sigma_* \over 0.3 ~{\rm km~s}^{-1} }\right)^2
\left({\mu_0 \over 10 M_\odot~{\rm pc}^{-1}  }\right)^{-1}
\end{equation} 
where we have  inserted the estimated velocity dispersion of the cluster \citep{oden02}
and a scaling value for the linear density.  We can see that
$q$ may be of order 1 for Palomar 5's tail just by looking at the above
scaling values.

The position of the first clump  gives us an estimate of the growth timescale
for the instability.   
 The first clump is visible only about 2 degrees away from the cluster center. This corresponds to 800 pc or a time 1.3~Gyr using the above drift velocity.  This implies
that the growth rate is approximately $\gamma \sim 0.7$~Gyr$^{-1}$ (assuming a single exponential growth timescale to the first clump).

The growth rate is related to the longitudinal velocity dispersion in the tail so
the estimated growth rate gives an estimate for the stellar velocity dispersion in the tail. 
For $q\approx 0.9$ we expect that the growth rate has
$\gamma {r_0 \over \sigma_*}=0.2$ (using Figure 1)
consequently we can estimate
\begin{equation}
\sigma_* \sim {\gamma r_0 \over 0.2} \sim 0.2 {\rm km~s}^{-1}
\end{equation}  
This dispersion and our previous estimate for $q$ implies that the
mass density in the tail is about $\mu \sim 4 M_\odot$pc$^{-1}$.  This is in between 
the density estimated from the mass loss rate,  $\mu\sim 8 M_\odot$pc$^{-1}$, and
that estimated from star counts $\mu \sim 2-3 M_\odot$pc$^{-1}$.
Had we used a lower value of the constant inside the absolute value in equation 
\ref{eqn:grav} we would have estimate a slightly higher growth grate and linear density.

The velocity dispersion we estimate above is significantly lower than the 2--4 km/s measured by \citet{oden09}.
We note that if the dispersion is isotropic a velocity dispersion of 1 km/s would cause
the tidal stream to spread at a rate of about 1~kpc~Gyr$^{-1}$.  This would exceed the width of the observed tail
a few degrees from the cluster (taking into account the estimated drift rate).
Consequently a velocity dispersion in the tail is expected to be similar to that estimated for the cluster or $\sim 0.3$ km/s \citep{oden02} and so consistent with our estimate.
Numerical simulations by \citet{dehnen04} have also concluded that the tail is ``cold.'' 
\citet{kupper10} suggest that stars escape the cluster from Lagrange points and
so the tail could have velocity dispersion lower than the cluster dispersion.
In short we find that the tail density and velocity
dispersion could be consistent with an interpretation of clumps in Palomar 5 as the fastest growing Jeans unstable mode of a stellar cylinder.    Here we used a distance
of 2 degrees between clumps to estimate the wavelength of the fastest growing mode but there could also be structure on 1 degree scales (see Figure 14 by \citealt{oden03}).
We used a 2 degree scale because our dispersion relation is appropriate for $kr_0>1$. 
Future work could improve calculation of the dispersion relation for the $k r_0 \sim 1 $ regime. 

\section{Estimating the tail density during tidal stripping}

The mass loss rate due to tidal disruption depends on the radial density
distribution of the cluster and on the Galactic tidal field.   Thus tidal evolution
models should be able to predict the linear mass density and velocity dispersion
in tidal tails.  To  estimate these quantities we explore the simplest model, that where
both background galaxy and cluster are described as spherically symmetric isothermal spheres  with dispersions $\sigma_g$ and $\sigma_d$, respectively.  It may also be useful to 
write things in terms of velocities of particles in circular orbits 
$v_g = \sqrt{2} \sigma_g$.
A balance of tidal force versus self gravity gives a relation for the tidal 
radius $r_t$ of the cluster
\begin{equation}
{r_t \over R} = {\sigma_d \over \sigma_g}
\end{equation}
where the cluster is at a radial distance $R$ from the galaxy center.
This gives a mass loss rate for the cluster
\begin{equation}
\dot M = {2 \sigma_d^2 \over G} {\sigma_d \over \sigma_g} \dot R
 \end{equation} 
 Material stripped from the cluster at the tidal radius will be on orbits that move away
 from the cluster because they have a different angular momentum.
 We let $\dot R = b v_g$ with unit-less parameter $b$ that depends on the cluster orbit
 with $b=0$ if on a circular orbit and $b$ of order unity for a radial orbit.
 Using equation (\ref{eqn:mu}) and (\ref{eqn:dzdt2}) and the above relation for $\dot M$,
  we can estimate the mass density in the tidal tails as
 \begin{equation}
 \mu = {dM \over dz} = \dot M {dt \over dz} = {2 \sigma_d^2 \over G} b
 \end{equation}
 
The dispersion in the tail should be similar to the velocity dispersion of the cluster, $\sigma_d$, and so using our estimate for $\mu$,
\begin{equation}
q \equiv { \sigma_d^2 \over 2G \mu} \approx {1 \over 4 b}.
\end{equation}
The above value for $q$ is of order unity for a moderately eccentricity orbit.
We expect that streams from tidally disrupting isothermal spheres with a moderately eccentric orbit have $q$ of order 1
and so are likely to have fastest growing mode with wavelength of order a few times the radius of the stream.  In this regime the growth rate can either
be estimated with a velocity $\sqrt{G\mu}$ or a velocity $\sigma_d$.

The estimates in this section are for an isothermal sphere which has radial density profile $\rho \propto r^{-2}$.  This can be compared to the core of a King model that has constant density.   The concentration of the cluster can be described by
a concentration parameter $c_t \equiv \log_{10}(r_t/R_0)$ where $R_0$ is the core radius and $r_t$
the tidal radius.  This parameter is used to describe King models. Exterior to $R_0$, clusters with high concentration parameter have density dropping more steeply with radius than those with low concentration parameter. If the density drops less steeply with radius than an isothermal sphere, as would be true for a cluster with low concentration parameter, then we would expect that the linear mass density in the tidal tail is higher than predicted from an isothermal sphere.  This in turn we expect would lower the $q$ of the tail, allowing shorter wavelength modes to grow.    If the cluster or galaxy is centrally concentrated then $q$ would be high, only very long wavelength modes would grow, and they would grow very slowly.

In summary we find that tidal streams from disrupting isothermal spheres have $q$ of order 1, with lower and higher concentration disrupting objects having tails with higher values and lower values of $q$, respectively.   We expect the separation between clumps to be set by the fastest growing wavelength which we expect should be of order a few times the tidal radius of the tidally disrupting cluster.  If the cluster is small then the timescale for stars to move away from the cluster or galaxy is long.  But in this case the instability takes longer to grow because the density is lower.   The two effects cancel out to leave an e-folding length scale for the growth of the instability in the tidal-tail that is only dependent on the tidal radius.

\section{Summary and Conclusion}

We have investigated Jeans instability of a stellar cylinder in the context of tidal tails.  As had previous works found (e.g., \citealt{fridman84}), we find there are always modes that can grow.   The wavelength and growth rate of the fastest growing mode depends on a $q$ parameter $q = {\sigma_*^2 \over 2G\mu_0} = 2/(k_J r_0)^2$ that is analogous
to the Toomre Q parameter.  Here $\mu$ is the linear mass density, $\sigma_*$ is
the longitudinal velocity dispersion, $r_0$ is the tail radius and $k_J$ the Jeans wave-number.   For $q\sim 1$, the fastest growing mode
has wavelength a few times the diameter of the tail and  growth rate  approximately equal to the tail velocity dispersion divided by the tail diameter.  Because the wavelength and timescale of maximum growth rate depend on the $q$ value and velocity dispersion in the tail, both can be roughly estimated from the spacing of observed clumps in the stream and the location where they appear.    

For a cluster or galaxy that is approximately an isothermal sphere we find that the disrupting system will have tidal tails with  $q$ of order 1.
Clusters or galaxies with low concentration parameters and on eccentric orbits could have tidal tails with lower $q$ parameters and so their modes of maximum growth rate should have shorter wavelengths and faster growth rates.
More concentrated objects would have tails for which the mode of maximum
growth rate is more slowly growing and longer wavelength than for low density or low concentration parameter objects.  All tidal tails could exhibit Jean instability, however if the cluster or galaxy is sufficiently concentrated then the wavelength of the fastest
growing mode in the tail would be large and
growth rate sufficiently long that the instability may be difficult to detect.  
The opposite is true for objects that are close to being completely disrupted;  Jeans
instability should be particularly prominent.

The tidal tails of Palomar 5 exhibit clumps with spacing approximately a few
times the tail width.   From this we have estimated a tail density and find it  approximately consistent with observational estimates.  This suggests
that the clumps in Palomar 5's tails are consistent with the fastest growing Jeans unstable mode.  Our estimated velocity dispersion is lower than measured by 
\citet{oden09} but consistent with the estimated cluster velocity dispersion.  
Palomar 5 has a low concentration parameter and so its tails would be expected
to exhibit a quickly growing, short wavelength, fastest growing Jeans unstable mode. 

We find that tidal tails from disrupting clusters or dwarf galaxies should display gravitational instabilities and from measurements of the wavelengths of the fastest growing mode, properties of the tail can be directly measured in a fashion only weakly dependent on the orbit or host galaxy dark matter profile.    More detailed understanding of the instability may yield constraints on the orbit and tidal mass loss rate that are complimentary to those from studies of shape of the tail on the sky or its velocity gradient.   

Here we have ignored the nature of the stellar orbits  in the tails as they move through the galaxy.   However density variations will be induced in the tail by its orbit.  
For Palomar 5 the growth timescale for the instability exceeds the orbital period so it may be sufficient to consider the tail density averaged over the orbit.   In other systems the orbit period may be shorter than the growth timescale for the fastest growing mode.  In this case at orbit apocenters a linear density increase could cause shorter wavelength modes to grow.  In both cases as the tail expands and density drops, shorter wavelength unstable modes may become stable.   

The Jeans instability model for clumps in tidal tails differs from that 
proposed by \citet{kupper08} who interpreted clumps in the tidal tails of Palomar 5 in terms of epicyclic over-densities.  By studying systems of different densities and with different orbits, future studies should be able to differentiate between the possible explanations for clumping  in stellar tidal tails.     Future work will determine if the clumping seen in N-body simulations of higher density tidal tails from dwarf galaxies \citep{justin10} is due to Jeans instability.

We thank Joss Bland-Hawthorn for helpful comments. Support for this work was provided by NSF through award AST-0907841.
Support for this work was provided by the National Aeronautics and Space Administration through Chandra Award Number Cycle 10-800424,  and Chandra Cycle 11-GO0-11012C  issued by the Chandra X-ray Observatory Center, which is operated by the Smithsonian Astrophysical Observatory for and on behalf of the National Aeronautics Space Administration under contract NAS8-03060. 
Support for this work, part of the Spitzer Space Telescope Theoretical Research Program, was provided by NASA through a contract issued by the Jet Propulsion Laboratory, California Institute of Technology under a contract with NASA.  

{}

\begin{thebibliography}{}

\bibitem[Chandrasekhar \& Fermi(1953)]{chandra53}
Chandrasekhar, S., \& Fermi, E.	
1953, ApJ, 118, 116	

\bibitem[Comprretta \& Quillen (2010)]{justin10}
Comparetta, J. \& Quillen, A. C. 2010, in preparation

 \bibitem[Dehnen et al.(2004)]{dehnen04}	  
Dehnen, W., Odenkirchen, M., Grebel, E. K., \& Rix, H.-W. 2004, AJ, 127, 2753	
	
\bibitem[Duc et al.(2000)]{duc00}
Duc, P.-A., Brinks, E., Springel, V., Pichardo, B., Weilbacher, P., 
Mirabel, I. F. 2000, AJ, 120, 1238

\bibitem[Elmegreen \& Efremov(1996)]{elmegreen96}
Elmegreen, B. G., \& Efremov, Y. N. 1996, ApJ, 466, 802	


\bibitem[Elmegreen \& Elmegreen(1983)]{elmegreen83}
Elmegreen, B. G., \& Elmegreen, D. M. 1983, MNRAS, 203, 31	


\bibitem[Elmegreen(1994)]{elmegreen94}
Elmegreen, B. G. 1994, ApJ, 433, 39	

\bibitem[Elmegreen(1979)]{elmegreen79}
Elmegreen, B. G. 1979, ApJ, 231, 372	


\bibitem[Fridman \& Polyachenko(1984)]{fridman84}
Fridman, A. M, \& Polyachenko, V. L., 1984, ``Physics of Gravitating Systems,'' Volume 1, (New York: Springer-Verlag)
 
\bibitem[Gnedin et al.(1999)]{gnedin99}
Gnedin, O. Y., Lee, H. M., \& Ostriker, J. P.	
1999, ApJ, 522, 935	

\bibitem[Grillmair \& Dionatos(2006)]{grillmair06}
Grillmair, C. J., Dionatos, O. 2006, ApJ, 641, L37	

\bibitem[Harris(1996)]{harris96}
Harris, W. E. 1996, AJ, 112, 1487

\bibitem[Johnston et al.(1999)]{johnston99}		
Johnston, K. V., Sigurdsson, S., \& Hernquist, L. 1999, MNRAS, 302, 771		

\bibitem[Just et al.(2009)]{just09}
Just, A., Berczik, P., Petrov, M. I., \& Ernst, A.	2009, MNRAS, 392, 969

\bibitem[Koch et al.(2004)]{koch04}
Koch, A., Grebel, E. K., Odenkirchen, M., Mart'nez-Delgado, D., \& Caldwell, J. A. R. 
2004, AJ, 128, 2274	
		
\bibitem[K\"upper et al.(2008)]{kupper08}
K\"upper, A. H. W., MacLeod, A., \& Heggie, D. C. 2008, MNRAS, 387, 1248

\bibitem[K\"upper et al.(2010)]{kupper10}
K\"upper, A. H. W.; Kroupa, P., Baumgardt, H., Heggie, D. C.
2010, MNRAS, 401, 105

\bibitem[Mayer et al.(2002)]{mayer02}
Mayer, L., Moore, B., Quinn, T., Governato, F., \& Stadel, J.
2002, MNRAS, 336, 119

\bibitem[Odenkirchen et al.(2001)]{oden01}
Odenkirchen, M. et al.   2001, ApJ, 548, L165	
	
\bibitem[Odenkirchen et al.(2002)]{oden02}
Odenkirchen, M., Grebel, E. K., Dehnen, W., Rix, H.-W., Cudworth, K. M.
2002, AJ, 124, 1497	
		
\bibitem[Odenkirchen et al.(2003)]{oden03}
Odenkirchen, M., Grebel, E. K., Dehnen, W., Rix, H.-W., Yanny, B., Newberg, H. J., Rockosi, C. M., Mart'nez-Delgado, D., Brinkmann, J., \& Pier, J. R.	
2003, AJ, 126, 2385	

\bibitem[Odenkirchen et al.(2009)]{oden09}
Odenkirchen, M., Grebel, E. K., Kayser, A., Rix, H.-W. \& Dehnen, W.
2009, AJ, 137, 3378

\bibitem[Ostriker(1964)]{ostriker64}
Ostriker, J. 1964, ApJ 140, 1529

\bibitem[Smith et al.(2010)]{smith10}
Smith, B. J., Giroux, M. L., Struck, C., Hancock, M., \& Hurlock, S.
2010, arXiv1001.0989, ApJS in press?


\end{thebibliography}
\end{document}